\RequirePackage{lineno}
\documentclass[10pt, superscriptaddress, a4paper, twocolumn,nofootinbib]{revtex4}
\usepackage[brazil]{babel}
\usepackage{epstopdf}
\usepackage{gensymb}
\usepackage{indentfirst}
\usepackage{graphicx}
\usepackage{amsmath,amssymb,amsfonts}
\usepackage{longtable,booktabs}
\usepackage[flushleft]{threeparttable}
\usepackage{breakurl}
\usepackage{tabularx,epsfig}
\usepackage{lscape}
\usepackage{rotating}
\usepackage[titles]{tocloft}

\renewcommand{\thesection}{\arabic{section}{.}}

\makeatletter
\@ifundefined{textcolor}{}
{%
 \definecolor{BLACK}{gray}{0}
 \definecolor{WHITE}{gray}{1}
 \definecolor{RED}{rgb}{1,0,0}
 \definecolor{GREEN}{rgb}{0,1,0}
 \definecolor{BLUE}{rgb}{0,0,1}
 \definecolor{CYAN}{cmyk}{1,0,0,0}
 \definecolor{MAGENTA}{cmyk}{0,1,0,0}
 \definecolor{YELLOW}{cmyk}{0,0,1,0}
 }



\usepackage{babel}
\usepackage[urlcolor=white]{hyperref}
\hypersetup{
    linkcolor=black,
    filecolor=black,      
    urlcolor=black,
}

\makeatother

\usepackage{babel}

\makeatother

\usepackage{babel}

\usepackage{titlesec}
\titleformat{\section}
    [block]{\normalfont\bfseries\large}{\rlap{\thesection}{}}{0em}
    {\hspace{.05\linewidth}\begin{minipage}[t]{.95\textwidth}}[\end{minipage}]

\def\C{\c{c}}

\begin{document}

\title{Uma andorinha s\'o n\~ao faz ver\~ao: 160 anos do legado de Richard Carrington

\vspace{0.3cm}

\footnotesize (One swallow does not make a summer: 160 years of Richard Carrington's legacy)}

\author{D. M. Oliveira}
\email{denny.m.deoliveira@nasa.gov}
\address{NASA Goddard Space Flight Center, Greenbelt MD, United States}
\address{Goddard Planetary Heliophysics Institute, University of Maryland Baltimore County, Baltimore, MD United States}

\vspace{0.5cm}

\begin{abstract}

	No dia 1$^o$ de setembro de 1859, o astr\^onomo brit\^anico Richard Carrington observou um comportamento an\^omalo no Sol. Aproximadamente 17 horas depois, efeitos magn\'eticos de escalas globais foram observados na Terra: auroras boreais e austrais em regi\~oes de baixas latitudes, e falhas em equipamentos telegr\'aficos na Europa e Am\'erica do Norte. Hoje, 160 anos depois, sabemos que essa conex\~ao solar-terrestre controla a atividade magn\'etica no espa\C{}o pr\'oximo da Terra e sua superf\'icie, e seus efeitos s\~ao objetos de uma disciplina denominada {\it clima espacial}. O objetivo principal deste modesto trabalho \'e apresentar brevemente ao leitor as observa{\C}\~oes de Carrington e as descobertas que levaram \`a conex\~ao entre fen\^omenos magn\'eticos solares e terrestres. Tamb\'em s\~ao brevemente discutidas as implica{\C}\~oes desta descoberta ao clima espacial com \^enfase na import\^ancia da prote\C{}\~ao de aparelhos tecnol\'ogicos presentes no geoespa{\C}o e no solo, al\'em de uma breve discuss\~ao do futuro da disciplina nas pr\'oximas d\'ecadas. \par

	\noindent {\footnotesize {\bf Palavras-chave:} Evento de Carrington, Intera\c{c}\~oes Sol-Terra, Clima espacial.}

	\vspace{0.30cm}

	On 1 September 1859, the British astronomer Richard Carrington observed an anomalous behavior on the Sun. Approximately 17 hours later, magnetic effects of global scales were observed at Earth: aurora borealis and australis in low-latitude regions, and telegraphic equipment failures in Europe and North America. Nowadays, 160 years later, we know that solar-terrestrial connections control the magnetic activity in the geospace and on the ground, and their effects are subject of a discipline named {\it space weather}. The main goal of this modest work is to briefly present to the reader Carrington's observations and the discoveries that led to the connection between solar and terrestrial magnetic phenomena. Implications of this discovery to space weather with emphasis on the protection of technological systems in the geospace and on the ground, and a brief discussion on the future of the discipline, are briefly presented as well. \par

	\\
	\\
	\\
	{Published in Revista Brasileira de Ensino de F\'isica, (2020), 42(1),\\ http://dx.doi.org/10.1590/1806-9126-RBEF-2019-0213}

	\vspace{0.40cm}

\end{abstract}

\def\C{\c{c}}

\maketitle 

	\section{O astr\^onomo Richard Carrington}
			 
		Richard Christopher Carrington (1826-1875) foi um astr\^onomo brit\^anico amador interessado em observa{\C}\~oes solares. Carrington observava diariamente a din\^amica da superf\'icie solar, tomando notas e armazenando dados em detalhes minuciosos. Com o resultado de sua an\'alise de manchas solares \cite{Eddy1976,Echer2003c,Vaquero2007b} ele descobriu que tais manchas executam movimentos latitudinais e a contagem destas apresenta n\'umeros m\'aximos com per\'iodos de aproximadamente 11 anos, o que ficou posteriormente conhecido como ciclo de atividade solar, ou simplesmente ciclo solar. Infelizmente, devido a problemas pessoais e familiares, Carrington n\~ao teve a oportunidade de publicar seu trabalho e apresentar seus resultados. A descoberta do ciclo solar \'e creditada ao astr\^onomo alem\~ao Gustav Sp\"{o}rer (1822-1895), que executou suas observa{\C}\~oes independentemente de Carrington. Sp\"{o}rer tamb\'em descobriu a lei que leva seu nome, que consiste na observa{\C}\~ao do aparecimento de novas manchas solares em extens\~oes latitudinais do equador solar, entre 30$^\circ$ e 40$^\circ$ dos dois hemisférios, que diminuem com o avan{\C}o do ciclo solar \cite{Maunder1890,Eddy1976,Vaquero2007b}. Devido a sua forma peculiar, os gr\'aficos dos resultados dessas observa{\C}\~oes s\~ao denominados diagramas borboletas \cite{Russell2016,Morris2019}. Entretanto, tal fen\^omeno foi primeiramente observado por Carrington. O livro de Stuart Clark conta a hist\'oria dos fatos que culminaram na trag\'edia que consistiu a vida de Richard Carrington \cite{Clark2007a}. \par

		Em geral, Richard Carrington n\~ao \'e conhecido na comunidade cient\'ifica atual somente pelo n\~ao reconhecimento de seus trabalhos em sua \'epoca, mas tamb\'em \'e conhecido por ser o primeiro a observar claramente uma erup{\C}\~ao solar e associ\'a-la, de maneira cuidadosa, a fen\^omenos magn\'eticos terrestres. Curiosamente, o evento de Carrington, como \'e conhecido hoje, n\~ao \'e somente a primeira erup{\C}\~ao solar a ser observada, mas \'e tamb\'em a erup{\C}\~ao solar com as consequ\^encias mais intensas ao ambiente terrestre observadas e registradas na hist\'oria. O objetivo deste trabalho \'e revisar as descobertas que levaram \`a conex\~ao entre as atividades magn\'eticas solar e terrestre, e reconhecer a import\^ancia do evento de Carrington no avan{\C}o do entendimento desta conex\~ao e seu impacto na disciplina denominada {clima espacial}. \par

	\section{Conex\~oes entre efeitos magn\'eticos\\solares e terrestres s\~ao estabelecidas}

		As manchas solares foram descobertas por Galileu Galilei (1564-1642) atrav\'es de observa{\C}\~oes por telesc\'opios no in\'icio do s\'eculo XVII \cite{Galilei1610}. Entretanto, houve pouco progresso nas observa{\C}\~oes de manchas solares no s\'eculo seguinte, com exce{\C}\~ao \`as observa{\C}\~oes do astr\^onomo brit\^anico Alexander Wilson (1714-1786) que associou as manchas solares\footnote{Hoje, sabemos que manchas solares s\~ao regi\~oes de temperaturas superficiais reduzidas relativamente ao seu entorno, causadas por fluxos magn\'eticos intensos, que inibem o processo de convec\C\~ao de plasma (por isso o car\'ater escuro das manchas). Explos\~oes solares ocorrem devido a um processo denominado {\it reconex\~ao magn\'etica} que por sua vez lan{\C}am grandes quantidades do plasma solar com energia cin\'etica intensa no espa{\C}o interplanet\'ario \cite{Russell2016}.} a ``buracos" na superf\'icie do sol \cite{Wilson1774}. Existem dados de manchas solares colhidos durante o s\'eculo XVIII, mas estes dados n\~ao s\~ao confi\'aveis devido sua esporacidade e m\'etodos imprecisos utilizados em sua coleta \cite{Eddy1976,Vaquero2007b,Hathaway2015}. \par

		O estudo de manchas solares teve um impulso consider\'avel gra{\C}as ao trabalho do farmac\'ologo alem\~ao Samuel Heinrich Schwabe (1789-1875), cujos resultados foram divulgados na revista {\it Astronomische Nachrichten} \cite{Schwabe1844}, uma das primeiras revistas cient\'ificas dedicadas \`a astronomia. Em seu trabalho, Schwabe identificou um ciclo de 10 anos associado ao n\'umero de manchas solares. Ap\'os refinadas observa{\C}\~oes executadas por Rudolf Wolfe (1816-1893), o per\'iodo do ciclo solar foi corrigido para o valor mais real\'istico de 11 anos \cite{Wolfe1852,Cliver2006}, que consiste na altern\^ancia entre dois m\'aximos ou m\'inimos consecutivos. O trabalho de Schwabe ganhou import\^ancia ap\'os ser popularizado pelo famoso cientista alem\~ao Alexander von Humboldt em seu livro {\it Cosmos} \cite{Humboldt1852}. \par

		O astr\^onomo e militar brit\^anico Edward Sabine (1788-1883) teve um papel importante na execu{\C}\~ao das Cruzadas Magn\'eticas, financiadas pela Coroa Brit\^anica na tentativa de estudar o campo magn\'etico terrestre \cite{Cawood1979}. Por volta do in\'icio do s\'eculo XIX, era amplamente conhecido que o campo magn\'etico da Terra era altamente vari\'avel, e tais varia\C\~oes interferiam na orienta\C\~ao de agulhas magn\'eticas, produzindo, por exemplo, grande impacto nos instrumentos de navega\C\~ao \cite{Cawood1979,Cliver2006,Clark2007a}. Ao analisar esses dados, Sabine descobriu que as varia{\C}\~oes magn\'eticas associadas com movimentos de agulhas magn\'eticas coincidiam estritamente com os n\'umeros de manchas solares descobertos por Schwabe e refinados por Wolfe \cite{Sabine1852}. Como resultado, uma eventual conex\~ao entre fen\^omenos solares e magnetismo terrestre foi claramente identificada pela primeira vez na hist\'oria da f\'isica solar. Entretanto, uma andorinha solit\'aria era necess\'aria para anunciar a chegada do ver\~ao. \par

		\begin{figure}
			\centering
			\includegraphics[width=0.48\textwidth]{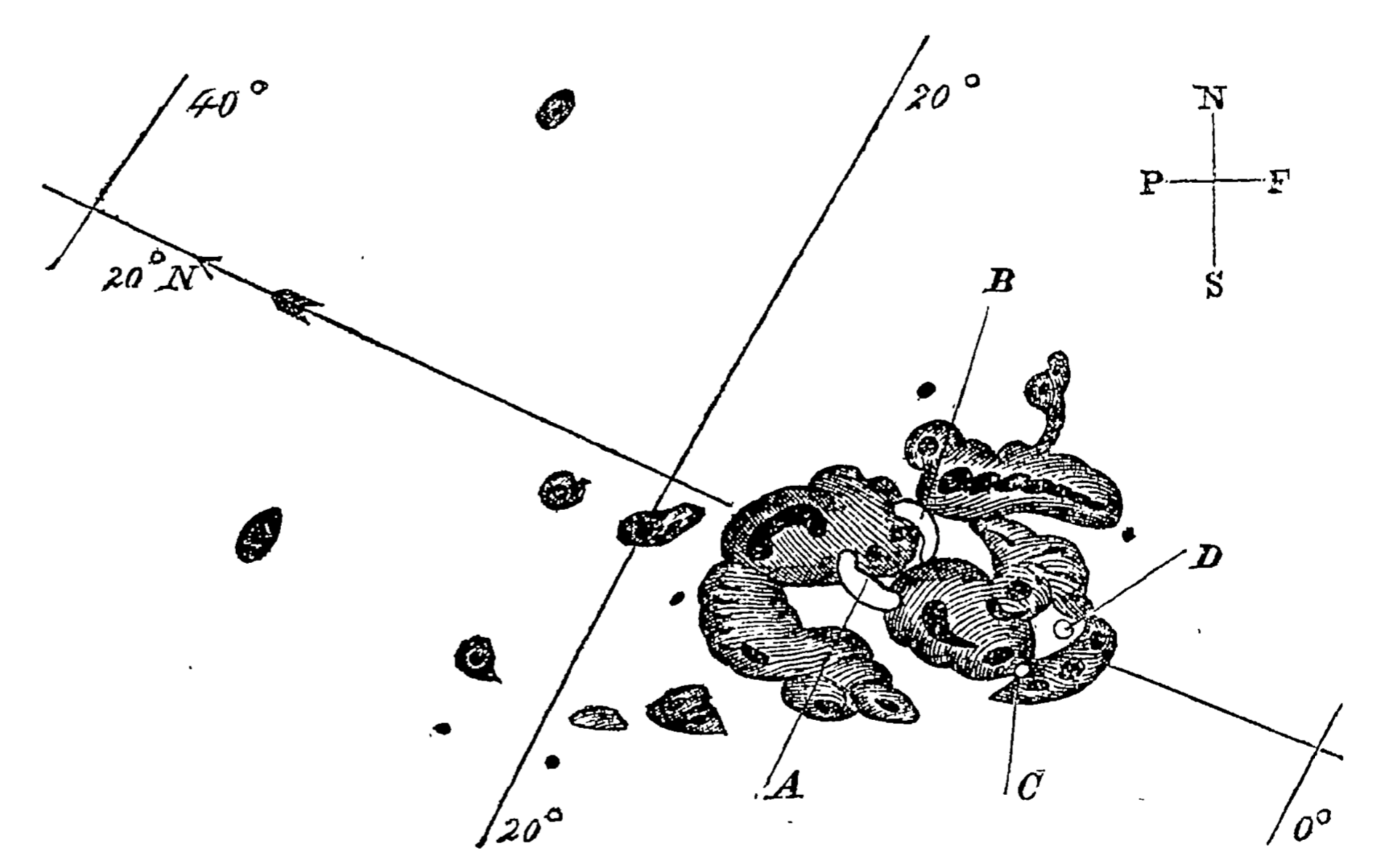}
			\caption{Diagrama esquem\'atico produzido por Carrington mostrando as massivas manchas solares observadas por ele em 1$^o$ de setembro de 1859. Extra\'ido da refer\^encia \cite{Carrington1859}.}
			\label{carrington_sunspots}
		\end{figure}

		No dia 1$^\circ$ de setembro de 1859, Carrington observava o Sol com seu telesc\'opio em sua casa, na cidade de Redhill, nos sub\'urbios de Londres, quando percebeu uma atividade diferente na superf\'icie solar. Carrington observou uma emiss\~ao solar extremamente brilhante que hoje \'e denominada erup{\C}\~ao solar. Maravilhado por sua observa{\C}\~ao e na \^ansia de encontrar uma outra pessoa para observar e testemunhar o fen\^omeno, Carrington deixou seu laborat\'orio momentaneamente. Quando voltou, para sua ``mortifica{\C}\~ao", o clar\~ao havia desaparecido. Gra{\C}as as suas habilidades observacionais, Carrington desenhou suas observa{\C}\~oes e enviou uma discuss\~ao dos resultados \`a revista {\it Monthly Notes of the Astronomical Society}, publicado em novembro de 1859 \cite{Carrington1859}. \par

		\begin{figure*}
			\centering
			\includegraphics[width=0.85\textwidth]{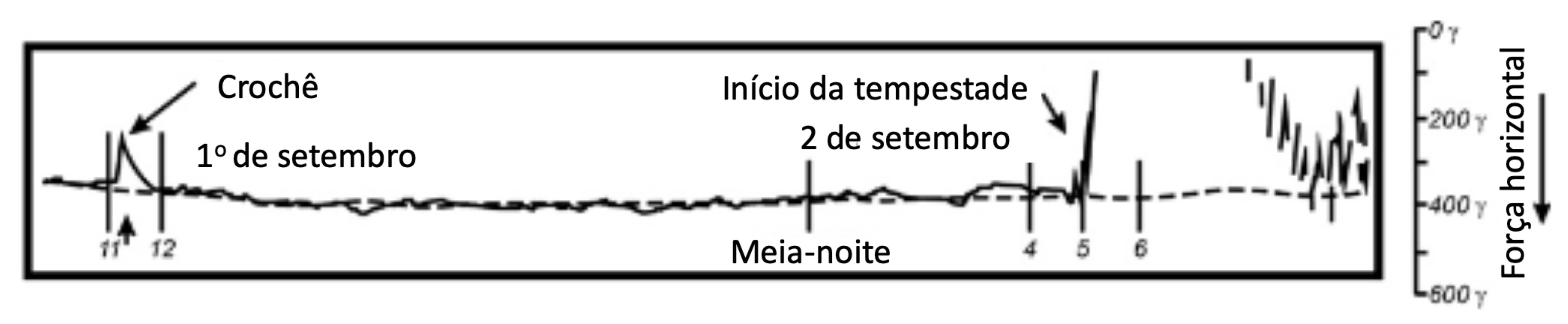}
			\caption{Perturba{\C}\~oes da componente horizontal do campo magn\'etico terrestre, ou a soma vetorial de suas componentes nas dire\C\~oes leste e norte, registradas por magnet\^ometros de solo localizados no observat\'orio magn\'etico Kew (coordenadas geogr\'aficas 51$^\circ$28' N, 359$^\circ$41.0' L) durante o evento de Carrington (1-2 de setembro de 1859). A tempestade magn\'etica occorreu devido \`a intensifica\C\~ao da corrente anelar e \`a redu\C\~ao severa da componente horizontal do campo magn\'etico terrestre \protect{\cite{Gonzalez1994,CostaJr2011b}}. Este diagrama foi publicado originalmente na refer\^encia \cite{Stewart1861}, e aqui adaptado da refer\^encia \cite{Cliver2006}.}
			\label{crochet}
		\end{figure*}

		O desenho publicado por Carrington \'e mostrado na Fig. \ref{carrington_sunspots}. A descri{\C}\~ao do evento, assim como da figura, \'e traduzida abaixo diretamente do artigo de Carrington \cite{Carrington1859} (tr\^es pontos entre colchetes indicam texto omitido pelo autor e textos entre colchetes foram adicionados pelo autor):

		\begin{quote}
			``Enquanto trabalhava na manh\~a de ter{\C}a-feira, 1$^o$ de setembro [de 1859], durante minhas observa{\C}\~oes costumeiras da forma e das posi{\C}\~oes das manchas solares, eu presenciei uma apari{\C}\~ao a qual acredito ser extremamente rara. A imagem do disco do Sol era, como sempre, projetada por mim [...] para produzir uma fotografia de aproximadamente 11 polegadas [27.94 cent\'imetros] de di\^ametro. Eu tinha diagramas confi\'aveis de todos os grupos de manchas solares [...] quando dentro da \'area do grande grupo ao norte [...] duas \'areas intensamente brancas e brilhantes surgiram, nas posi{\C}\~oes indicadas no diagrama anexado [Fig. \ref{carrington_sunspots}] pelas letras A e B, e das formas dos espa{\C}os deixados em branco. Minha primeira impress\~ao foi que por alguma chance um raio de luz tinha penetrado por um buraco na tela anexada ao objeto, pelo qual a imagem era direcionada \`a sombra, pois a luminosidade era muito semelhante \`a luz solar direta; mas, [...] eu presenciei um fen\^omeno muito diferente. Ent\~ao eu anotei a hora prontamente pelo cron\^ometro, e vi a explos\~ao aumentar rapidamente, [...], eu apressadamente corri para buscar algu\'em para testemunhar a exibi{\C}\~ao comigo, mas retornando dentro de 60 segundos, eu fui mortificado ao encontrar que [o clar\~ao] tinha mudado e enfraquecido. Muito brevemente depois os \'ultimos tra{\C}os se foram, e embora eu tenha permanecido assistindo [o aparato] estritamente por aproximadamente 1 hora, nada ocorreu novamente. Os \'ultimos tra{\C}os ocorreram em C e D, as \'areas que viajaram consideravelmente das suas primeiras posi{\C}\~oes e desaparecendo como dois pontos brilhantes de luz branca. A hora da primeira explos\~ao n\~ao foi diferente de 15 segundos de 11 h:18 min, hora local de Greenwich, e 11 h:23 min foi anotada como a hora do desaparecimento."
		\end{quote}

		De acordo com seu relato, Carrington observou a erup{\C}\~ao solar entre 11 h:18 min e 11 h:23 min do dia 1$^o$ de setembro de 1859. Balfour Stewart (1828-1887) reportou sobre a ocorr\^encia de uma perturba{\C}\~ao magn\'etica moderada registrada pelo observat\'orio de Kew, no sudoeste de Londres, ``o mais pr\'oximo de 11 h:15 min [hora local de Greenwich] poss\'ivel" \cite{Stewart1861}. Esta perturba\C\~ao \'e mostrada na Fig. \ref{crochet}. Hoje sabemos que essa perturba{\C}\~ao, denominada croch\^e\footnote{Esta perturba{\C}\~ao tem esse nome por apresentar, quando representada por um gr\'afico de s\'erie temporal, uma semelhan{\C}a com uma agulha de croch\^e.} magn\'etico, \'e uma perturba{\C}\~ao da camada superior ionizada da atmosfera terrestre (isto \'e, a ionosfera) causada por erup{\C}\~oes solares intensas \cite{Cliver2006}. Coincidentemente, outro astr\^onomo, Richard Hodgson (1804-1872) \cite{Hodgson1859}, tamb\'em observou, ao norte de Londres, a mesma erup\C\~ao solar em um intervalo de tempo coincidente com as observa\C\~oes reportadas por Carrington \cite{Carrington1859}. \par

	\section{A \lowercase{andorinha solit\'aria do ver\~ao de 1859}}

		O diagrama da perturba{\C}\~ao na componente horizontal do campo magn\'etico terrestre reportada por Stewart (Fig. \ref{crochet}) tamb\'em mostra uma outra perturba{\C}\~ao que ocorreu aproximadamente \`as 05:00 h (hora local de Greenwich), ou 17.6 h depois do croch\^e magn\'etico. A perturba{\C}\~ao foi t\~ao intensa que as medidas atingiram e ultrapasarram a escala m\'axima dos instrumentos, saturando as medi\C\~oes \cite{Stewart1861}. De fato, de acordo com o nosso conhecimento atual, os \'unicos instrumentos que forneceram medidas n\~ao saturadas durante aquele evento foram os instrumentos da esta{\C}\~ao magn\'etica Colaba, localizada em Bombaim (hoje Mumbai), \'India \cite{Tsurutani2003b}. \par

		\begin{figure*}
			\includegraphics[width=0.84\textwidth]{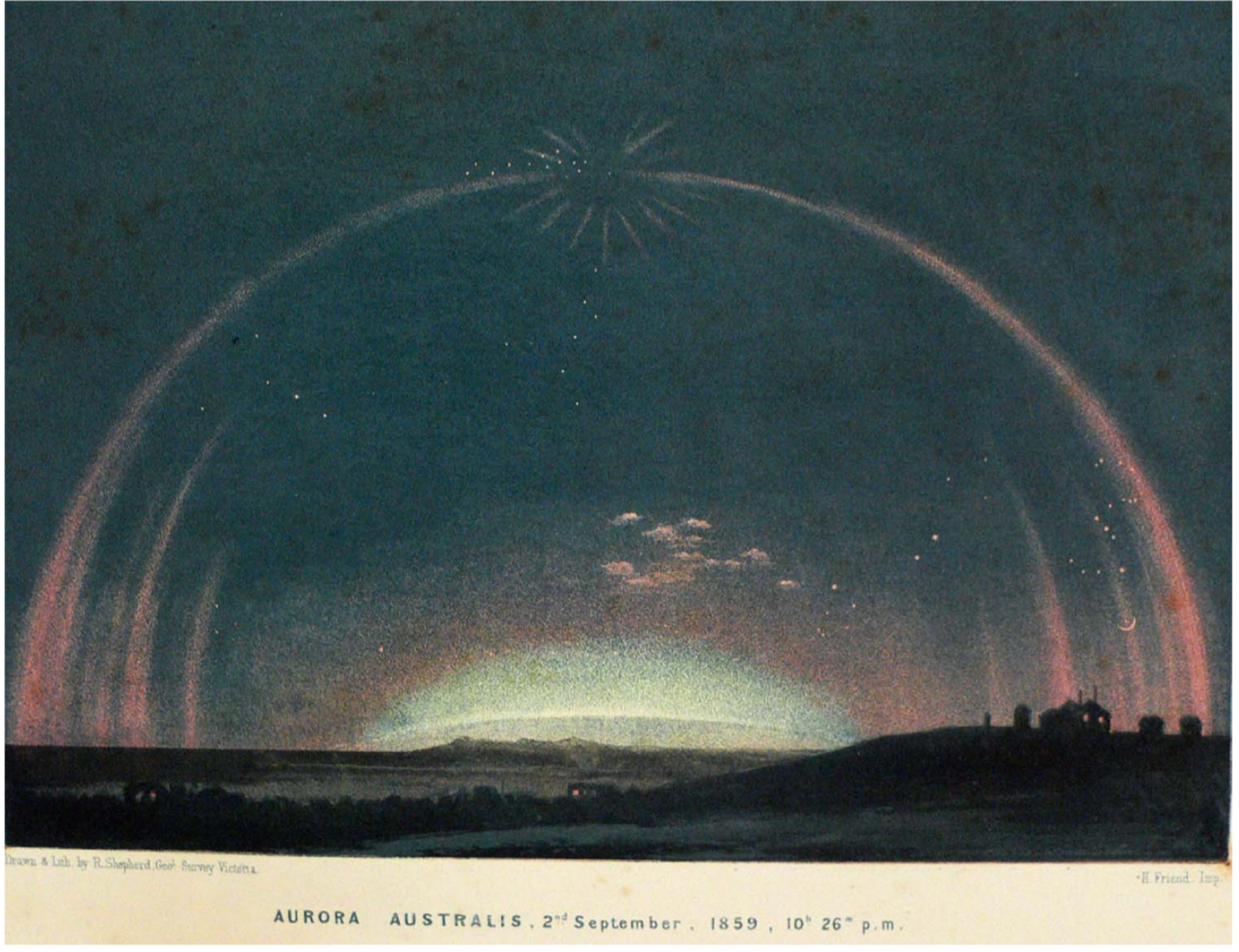}
			\caption{Diagrama de uma aurora austral observada em Melbourne, Austr\'alia, durante o evento de Carrington. Esta figura foi originalmente publicada na refer\^encia \cite{Neumayer1864}, e mostrada aqui como extra\'ida da refer\^encia \cite{Hayakawa2018b}.}
			\label{melbourne}
		\end{figure*}

		Os efeitos dessa perturba\C\~ao no campo magn\'etico terrestre, ou tempestade magn\'etica, foram rapidamente notados em escalas globais. Auroras boreais e austrais foram vistas em diversas regi\~oes de baixas latitudes (menores que 60$^\circ$), incluindo Hava\'i, Fl\'orida, e Washington, D.C. nos Estados Unidos \cite{Kimball1960,Green2006,Hayakawa2018b}, M\'exico \cite{Gonzalez-Sparza2018}, Col\^ombia \cite{Cardenas2016}, Espanha \cite{Farrona2011a}, Austr\'alia \cite{Humble2006}, \'Asia Oriental \cite{Hayakawa2016} e Chile, Portugal, leste da R\'ussia, Austr\'alia e Nova Zel\^andia \cite{Hayakawa2019b}. A Fig. \ref{melbourne} mostra um diagrama representando uma aurora austral vermelha observada na cidade de Melbourne, Austr\'alia, no dia 2 de setembro de 1859 \cite{Neumayer1864,Hayakawa2018b}. Testemunhas reportaram ter visto fios e equipamentos telegr\'aficos em chamas nos Estados Unidos, Canad\'a e Europa \cite{Loomis1861}. Em Washington, D.C., populares solicitaram a ajuda de bombeiros porque interpretaram uma aurora vermelha intensamente brilhante como um inc\^endio de grandes propor{\C}\~oes nos arredores da cidade \cite{Clark2007a,Odenwald2015}. Ainda na mesma cidade, um operador de tel\'egrafo faleceu em resultado de um choque el\'etrico proveniente das teclas do aparelho, e uma outra fatalidade semelhante tamb\'em ocorreu na cidade da Filad\'elfia, nos Estados Unidos \cite{Clark2007a}. Entretanto, pelo menos com base na aus\^encia de relatos da \'epoca, auroras austrais n\~ao foram observadas no territ\'orio brasileiro durante o evento de Carrington devido a latitudes magn\'eticas extremamente baixas do Brasil durante aquela \'epoca \cite{Hayakawa2019b}. \par

		Apesar dessas evid\^encias marcantes representadas pelas atividades magn\'eticas e seus resultados discutidos acima, Carrington foi muito cuidadoso e tentou evitar uma poss\'ivel conex\~ao entre erup\C\~oes solares e efeitos magn\'eticos terrestres. Richard Carrington escreveu em seu artigo {\it one swallow does not make a summer}, ou uma andorinha s\'o n\~ao faz ver\~ao \cite{Carrington1859}. Esse mist\'erio foi ent\~ao deixado para ser desvendado pelas futuras gera{\C}\~oes de f\'isicos solares e geof\'isicos.

	\section{Uma explica{\C}\~ao definitiva \'e fornecida}

		O progresso do entendimento das rela{\C}\~oes entre atividades magn\'eticas solares e terrestres foi relativamente lento durante o final do s\'eculo XIX e in\'icio do s\'eculo XX. Tal lentid\~ao tem duas explica{\C}\~oes: primeiro, os instrumentos de medi{\C}\~ao eram muito rudimentares e sat\'elites presentes no espa{\C}o interplanet\'ario inexistiam na \'epoca; segundo, devido a raridade de eventos intensos e a falta de uma amostra estat\'istica com um n\'umero razo\'avel de eventos, tal conex\~ao era frequentemente desafiada, como o fez Lord Kelvin (William Thonmson, 1824-1907). Lord Kelvin, baseado em c\'alculos de conserva{\C}\~ao de energia, concluiu que nosso Sol \'e incapaz de gerar energia para produzir uma tempestade magn\'etica \cite{Kelvin1893,Cliver2006}. Os c\'alculos de Lord Kelvin estavam corretos quando considerava o espa{\C}o interplanet\'ario completamente vazio. No entanto, ele ignorava a exist\^encia de um fluxo constante de part\'iculas eletricamente carregadas provenientes do Sol viajando com um campo magn\'etico aprisionado a elas, conhecido hoje como o vento solar \cite{CostaJr2011b,Echer2006b,Oliveira2016b}. Entretanto, esta conex\~ao come{\C}ou a ser mais aceita ap\'os os trabalhos de E. Walter Maunder (1851-1926), que encontrou uma rela{\C}\~ao recorrente entre um per\'iodo de aproximadamente 27 dias (uma rota{\C}\~ao de Carrington) e atividade magn\'etica terrestre \cite{Maunder1905}. \par

		\begin{figure*}
			\centering
			\includegraphics[width=0.79\textwidth]{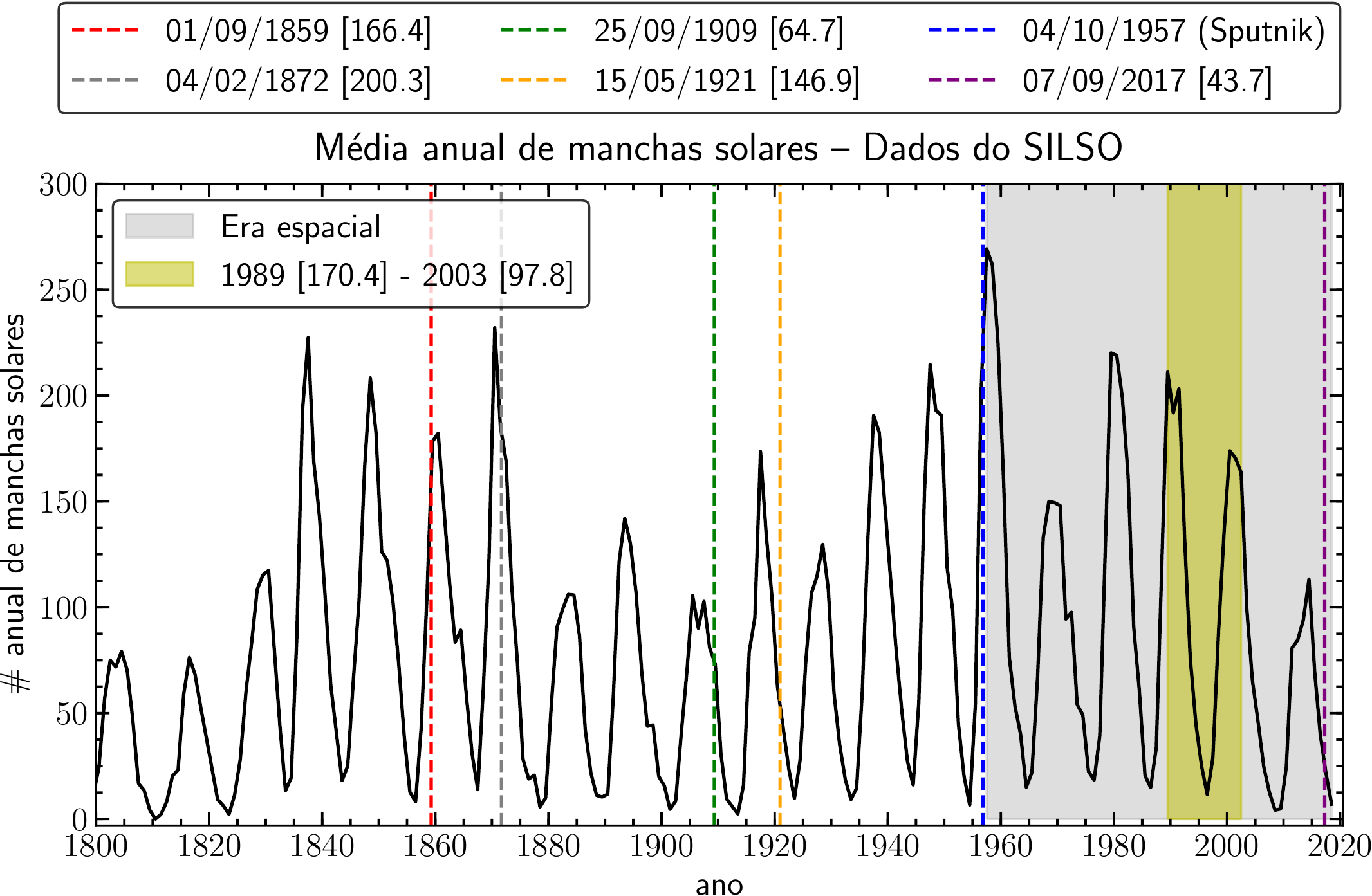}
			\caption{Gr\'afico mostrando m\'edias anuais de manchas solares no per\'iodo 1800 a 2018. Os dados foram obtidos do SILSO website, veja se\C\~ao 5 ({\it Sunspot Index and Long-term Solar Observations}). As linhas tracejadas mostram tempestades magn\'eticas conhecidas das quais alguns registros hist\'oricos t\^em sido encontrados recentemente. Os n\'umeros entre colchetes correspondem \`as m\'edias mensais de manchas solares registradas para o m\^es de ocorr\^encia da tempestade magn\'etica. A \'area hachurada em cinza corresponde ao intervalo conhecido como Era Espacial, que se iniciou com o lan{\C}amento do sat\'elite sovi\'etico {\it Sputnik}, enquanto que a \'area hachurada amarela mostra o intervalo entre a tempestade magn\'etica de mar{\C}o de 1989 e as tempestades magn\'eticas de 2003 que, por terem ocorrido no final do m\^es de outubro, foram denominadas como as tempestades de Halloween.}
			\label{silso_sunspots}
		\end{figure*}

		Os trabalhos do cientista noruegu\^es Kristian Birkeland (1867-1917) com os ``experimentos de terrela" contribu\'iram com o entendimento da natureza f\'isica do vento solar. Ao lan{\C}ar um feixe de el\'etrons na dire{\C}\~ao de uma esfera met\'alica magnetizada ({\it terrella}), Birkeland notou o aparecimento de estruturas parecidas com auroras boreais e austrais em ambos hemisf\'erios \cite{Birkeland1908}. Ele atribuiu esses efeitos \`a natureza de radia{\C}\~ao corpuscular do feixe de el\'etrons. O matem\'atico e geof\'isico brit\^anico Sydney Chapman (1888-1970) usou a hip\'otese corpuscular de Birkeland para construir uma teoria de tempestades magn\'eticas \cite{Chapman1919}, gerando o conceito da estrutura que hoje \'e conhecida como a magnetosfera terrestre \cite{Echer2010a,Russell2016,Oliveira2016b}. \par

		Com respeito ao evento de Carrington, duas perguntas ainda n\~ao tinham sido respondidas: o que foi causado na superf\'icie da Terra, e/ou no ambiente espacial no entorno desta pela erup{\C}\~ao solar observada por Carrington? O croch\^e magn\'etico, a perturba{\C}\~ao magn\'etica subsequente, ou ambos? Uma evid\^encia de que perturba{\C}\~oes magn\'eticas, cujas assinaturas s\~ao semelhantes ao croch\^e magn\'etico, podem ser causadas por erup\C\~oes solares foi fornecida pelo cientista americano J. Howard Dellinger (1886-1962), que explicou que interfer\^encias em ondas de r\'adio de curto comprimento de onda (10 cm -- 10 m) podem ocorrer coincidentemente com erup{\C}\~oes solares \cite{Dellinger1936,Cliver2006}. \par

		A resposta final \`a quest\~ao sobre o evento de Carrington foi finalmente fornecida pelo geof\'isico e estat\'istico alem\~ao Julius Bartels (1899-1964). Baseado nas descobertas de Dellinger \cite{Dellinger1936}, Bartels \cite{Bartels1937} conectou o croch\^e magn\'etico registrado pelo observat\'orio de Kew \cite{Stewart1861} \`a erup{\C}\~ao solar observada por Carrington \cite{Carrington1859}, e associou a ideia de emiss\~oes solares corpusculares de Birkeland \cite{Birkeland1908} e Chapman \cite{Chapman1919} \`a perturba{\C}\~ao geomagn\'etica intensa registrada pelo mesmo observat\'orio \cite{Stewart1861}. Portanto, o evento de Carrington era explicado corretamente pela primeira vez na hist\'oria da geof\'isica e heliof\'isica. Depois de duas d\'ecadas do trabalho de Bartels, a explora\C\~ao do geoespa{\C}o se iniciava.

	\section{Atividade solar e clima espacial}

		A Fig. \ref{silso_sunspots} mostra dados de m\'edias anuais de manchas solares dispon\'iveis no website SILSO ({\it Sunspot Index and Long-term Solar Observations}, \url{http://www.sidc.be/silso/datafiles}) \cite{Clette2016a,Clette2016b}. O gr\'afico (linha preta cont\'inua) mostra dados de manchas solares de 1800 at\'e 2018. \par

		A figura mostra claramente que o registro temporal do n\'umero de manchas solares apresenta per\'iodos de altern\^ancia entre m\'aximos e m\'inimos, cujo per\'iodo, de aproximadamente 11 anos, \'e definido como o ciclo solar. A figura tamb\'em mostra um intervalo com aproximadamente 19 ciclos. A linha tracejada vermelha da Fig. \ref{silso_sunspots} indica o evento de Carrington de 1$^o$ de setembro de 1859. A m\'edia mensal de manchas solares para aquele m\^es foi de 200.9. As caracter\'isticas da erup{\C}\~ao solar e da atividade magn\'etica terrestre subsequente, principalmente ao que diz respeito a auroras brilhantes e efeitos em equipamentos telegr\'aficos, t\^em sido amplamente discutidos na literatura \cite{Carrington1859,Loomis1861,Tsurutani2003b,Cliver2006,Clark2007a,Clark2007b,Hathaway2015,Hayakawa2019b}. \par

		A tempestade magn\'etica de 4 de fevereiro de 1872 tamb\'em causou visualiza\C\~oes deslumbrantes de auroras boreais e austrais em regi\~oes de baixas latitudes em \'areas diferentes do globo, al\'em de danos consider\'aveis em sistemas telegr\'aficos \cite{Meldrun1872,Silverman2008,Hayakawa2018a}. A posi{\C}\~ao desta tempestade magn\'etica na evolu{\C}\~ao temporal das manchas solares \'e mostrada na Fig. \ref{silso_sunspots} por uma linha tracejada cinza. A m\'edia mensal do n\'umero de manchas solares para fevereiro de 1872 \'e 200.3. A linha tracejada verde indica outra tempestade solar, no dia 25 de setembro de 1909, com m\'edia mensal de manchas solares de 64.7. Esta tempestade superintensa\footnote{Para mais detalhes sobre crit\'erios de categoriza\C\~ao da intensidade de tempestades magn\'eticas, veja, e.g., a refer\^encia \cite{Gonzalez1994}.} gerou auroras boreais e austrais observadas na \'Africa do Sul, Jap\~ao, Austr\'alia, Sudeste Asi\'atico e Oceano \'Indico \cite{Hayakawa2019a,Love2019a}. A tempestade magn\'etica de 15 de maio de 1921 (linha tracejada cor de laranja) causou s\'erios danos a sistemas telegr\'aficos, telef\^onicos, e ferrovi\'arios na cidade de Nova Iorque, nos Estados Unidos \cite{Love2019b}. A m\'edia mensal de manchas solares para o m\^es de maio de 1921 \'e 146.9. \par

		A regi\~ao destacada em cinza mostra a \'epoca denominada Era Espacial, que se estende do lan{\C}amento do sat\'elite sovi\'etico {\it Sputnik} (4 de outubro de 1957) at\'e os dias atuais. {\it Sputnik} tem uma import\^ancia hist\'orica porque, al\'em de ser um marco da aplica{\C}\~ao tecnol\'ogica na explora{\C}\~ao espacial e desenvolvimento educacional \cite{Wissehr2011}, o sat\'elite tamb\'em teve grande import\^ancia na corrida armament\'icia entre os Estados Unidos e a Uni\~ao Sovi\'etica durante o per\'iodo denominado Guerra Fria \cite{Peoples2008}. Dados do {\it Sputnik} foram utilizados pela primeira vez para associar deca\'imentos bruscos de sat\'elites de baixas \'orbitas \`a tempestades magn\'eticas \cite{Jacchia1959,Oliveira2017b}. Com respeito \`a atividade solar, curiosamente, os n\'umeros m\'aximos de manchas solares t\^em sido cada vez menores em cada ciclo solar na Era Espacial, com exce{\C}\~ao do n\'umero m\'aximo de manchas solares que ocorreu por volta de 1970. Como resultado, a atividade solar tem se reduzido enquanto a depend\^encia tecnol\'ogica da sociedade moderna tem aumentado continuamente nas \'ultimas d\'ecadas. \par 

		A faixa hachurada em amarelo mostra um per\'iodo solar altamente ativo entre 1989 e 2003, com os conhecidos eventos da tempestade magn\'etica de mar{\C}o de 1989, que causou a destrui{\C}\~ao de transformadores que levou ao blecaute em Quebec, no Canad\'a \cite{Allen1989}, e as conhecidas ``tempestades de Halloween" de outubro de 2003 \cite{Lopez2004}, que causaram o cancelamento e adiamento de diversos voos na Am\'erica do Norte \cite{NRC2008}. Como resultado de erup{\C}\~oes solares intensas, part\'iculas solares provenientes do vento solar \cite{CostaJr2011b,Echer2006b,Oliveira2016b} t\^em acesso a regi\~oes de altas latitudes das regi\~oes superiores da atmosfera terrestre. Tais part\'iculas altamente energ\'eticas ($>$500 MeV, or 1.0$\times$10$^{6}$ el\'ectron-volts) colidem mais frequentemente com as part\'iculas neutras da atmosfera que por sua vez interagem com avi\~oes, aumentando doses de radia{\C}\~ao em passageiros e tripula{\C}\~ao, particularmente em voos transpolares \cite{Schrijver2015}. Durante tempestades magn\'eticas intensas, varia{\C}\~oes de campos el\'etricos intensos na atmosfera terrestre ionizada geram correntes el\'etricas que, por sua vez, se acoplam a condutores artificiais no solo (linhas de transmiss\~ao de energia el\'etrica e transformadores), produzindo as denominadas correntes geomagn\'eticas induzidas (do ingl\^es {\it geomagnetically induced currents}, GICs) \cite{Viljanen1998,Pirjola2000,Schrijver2015}. Al\'em do blecaute em Quebec, GICs de alta intensidade causaram a completa destrui{\C}\~ao de um transformador de uma usina el\'etrica no estado de Nova Jersey, Estados Unidos, durante a tempestade magn\'etica de mar{\C}o de 1989 \cite{Kappenman2010,Oliveira2017d}. \par

		A linha tracejada roxa da Fig. \ref{silso_sunspots} mostra, dentro do per\'iodo analisado, a \'ultima tempestade magn\'etica de intensidade consider\'avel que ocorreu em 7 de setembro de 2017. Esta \'e uma das tempestades magn\'eticas mais intensas do ciclo solar atual, o mais fraco da Era Espacial. A m\'edia mensal de manchas solares durante o m\^es de setembro de 2017 foi a mais baixa (43.7) das tempestades magn\'eticas representadas na Fig. \ref{silso_sunspots}. Apesar disso, diversos efeitos de clima espacial t\^em sido associados \`aquela tempestade, como, por exemplo, o aumento consider\'avel do fluxo de part\'iulas carregadas no geospa{\C}o \cite{Bruno2019}, interfer\^encia em sinais de sistemas de sat\'elites de GPS \cite{Sato2019}, e efeitos de GICs em transmiss\~oes el\'etricas e oleodutos na Finl\^andia \cite{Dimmock2019}. \par

		Uma an\'alise cuidadosa da Fig. \ref{silso_sunspots} mostra que tempestades solares superintensas tendem a ocorrer em torno de n\'umeros m\'aximos de manchas solares, ou durante as fases de decl\'inio e ascens\~ao do n\'umero de manchas solares. Os n\'umeros anuais e mensais de manchas solares associados a essas tempestades s\~ao geralmente altos, com exce\C\~ao da tempestade magn\'etica de setembro de 2017. Entretanto, esta \'ultima tempestade n\~ao satisfaz condi\C\~oes bem estabelecidas para a classifica\C\~ao da intensidade de tempestades magn\'eticas como superintensas, como por exemplo a varia\C\~ao abrupta da componente horizontal do campo magn\'etico terrestre medida em solo \cite{Meng2019}. \par

		De acordo com a Fig. \ref{silso_sunspots}, as manchas solares apresentam outro comportamento peri\'odico peculiar. As amplitudes m\'aximas de cada ciclo solar, quando observadas em longo prazo (alguns ciclos solares), apresentam per\'iodos mais longos que correspondem a 77-88 anos, ou 7-8 ciclos solares regulares. Tal fen\^omeno \'e denominado {\it ciclo de Gleissberg} \cite{Gleissberg1967,Hathaway2015,Vazquez2016}, descrito claramente por Wolfgang Gleissberg (1903-1986) ap\'os an\'alises dos dados de manchas solares de Wolfe. H\'a evid\^encias de que esse padr\~ao tem se repetido nos \'ultimos 12000 anos, como mostram resultados de an\'alise de $^{14}$C das varia\C\~oes de is\'otopos cosmog\^enicos associados \`a atmosfera terrestre \cite{Peristykh2003}. Efeitos do ciclo de Gleissberg t\^em sido notados recentemente em registros hist\'oricos de auroras boreais no per\'iodo de 1600-2015 \cite{Vazquez2016}.

	\section{Conclus\~ao}

		Neste trabalho, as observa{\C}\~oes de Richard Carrington foram brevemente apresentadas e reconhecidas como a primeira tentativa, embora de forma extremamente cautelosa, em conectar fen\^omenos solares com os subsequentes fen\^omenos de atividade magn\'etica na Terra. Foi mostrado que ap\'os as observa{\C}\~oes de Carrington, a compreens\~ao de tal conex\~ao come{\C}ou a progredir, embora de forma modesta, gra{\C}as a descoberta da correla{\C}\~ao estreita entre atividade solar e atividade magn\'etica terrestre. Na primeira metade do s\'eculo XX, ao associar os resultados experimentais obtidos por Birkeland e Dellinger, juntamente com os resultados te\'oricos de Chapman, Bartels foi capaz de explicar o evento de Carrington pela primeira vez de maneira clara e conclusiva. De acordo com Bartels, o croch\^e magn\'etico foi causado pela erup{\C}\~ao solar (observada por Carrington), enquanto que a tempestade magn\'etica foi causada pela intera\C\~ao da ``radia{\C}\~ao corpuscular" (hoje conhecida como plasma) com a magnetosfera terrestre. O evento de Carrington foi um dos primeiros sinais da concep{\C}\~ao da disciplina de clima espacial, e claramente abriu o caminho para o desenvolvimento da disciplina, a qual decolou ap\'os o in\'icio da Era Espacial (1957 em diante) \cite{Cade2015}. A hist\'oria da explora{\C}\~ao do espa{\C}o interplanet\'ario n\~ao faz parte do escopo deste artigo, mas certamente \'e um assunto interessante para um trabalho futuro. \par

		Atualmente, o clima espacial e seus efeitos t\^em grande import\^ancia devido ao impacto que eles podem causar em nossa sociedade moderna e altamente tecnol\'ogica. Danos a sat\'elites de comunica{\C}\~oes e astronautas no geoespa{\C}o causados pelo aumento excessivo de radia{\C}\~ao, danos a redes e sistemas de transmiss\~ao el\'etrica causados por GICs, as quais tamb\'em causam corros\~ao de oleodutos, s\~ao alguns exemplos \cite{NRC2008,Schrijver2015}. Como resultado, o governo dos Estados Unidos tem recentemente reconhecido os efeitos de clima espacial como amea{\C}as naturais que podem causar s\'erios impactos econ\^omicos e sociais na sociedade contempor\^anea \cite{NSWS2015,NSWAP2015}. Esses relat\'orios tamb\'em enfatizam o estudo de condi\C\~oes extremas durante eventos de clima espacial, no qual o evento de Carrington desempenha um papel importante como evento base para an\'alise de dados e simula\C\~oes num\'ericas \cite{Ngwira2013b,Ngwira2014,Buzulukova2018}. A combina{\C}\~ao de an\'alises de dados fornecidos por plataformas espaciais localizadas na vizinhan\C{a} do nosso planeta, bem como de dados de instrumentos de solo aliadas a an\'alises de modelagens computacionais aprimora o entendimento de clima espacial e suas consequ\^encias atrav\'es de previs\~oes de tais eventos em curto, m\'edio e longo prazos \cite{Buzulukova2018}. \par

		De volta \`a Fig. \ref{silso_sunspots}, \'e poss\'ivel ver claramente dois ciclos de Gleissberg, cujos m\'inimos est\~ao nos intervalos aproximados de 1810-1885 e 1930-2010. Tendo em vista o ciclo de Gleissberg (e as manchas solares da Fig. \ref{silso_sunspots}), \'e razo\'avel argumentar que nas pr\'oximas d\'ecadas a atividade solar tem grande probabilidade de aumentar. Portanto, com a depend\^encia de tecnologias em n\'iveis ainda maiores, pode-se concluir que o estudo de clima espacial, suas aplica{\C}\~oes e previs\~oes ter\~ao grande import\^ancia e impacto no futuro de uma sociedade altamente tecnol\'ogica. Como resultado, n\~ao \'e dif\'icil imaginar que, assim como previs\~oes di\'arias do tempo (meteorologia), previs\~oes peri\'odicas de clima espacial marcar\~ao presen{\C}a constante nos meios de comunica{\C}\~ao do futuro.

	\section{Uma nota pessoal}

		De acordo com o website \url{www.academictree.org}, Julius Bartels ({\it Universit\"at G\"ottingen}, Alemanha) orientou o Ph.D. de Walter Kertz ({\it Universit\"at G\"ottingen}, Alemanha), que orientou o Ph.D. de Fritz Manfred Neubauer ({\it TU Braunschweig}, Alemanha), que orientou o Ph.D. de Joachim Raeder ({\it Universit\"{a}t zu K\"{o}ln}, Alemanha), que orientou o Ph.D. do autor deste trabalho na {\it University of New Hampshire}, Estados Unidos. A descoberta desta conex\~ao entre o autor deste trabalho e Julius Bartels, o cientista que claramente explicou a conex\~ao entre eventos magn\'eticos solares e terrestres, trouxe uma satisfa{\C}\~ao pessoal e especial ao autor.

	\section*{Agradecimentos}

		O autor agradece a NASA (ag\^encia espacial norte-americana) pelo apoio financeiro atrav\'es dos projetos 13-SRITM132-0011 e HSR-520 MAG142-0062 sob contrato com a UMBC. O autor tamb\'em agradece ao Dr. Marcos Silveira (NASA/GSFC) e ao professor Alexandre Emygdio (Col\'egio Exatus) pela cuidadosa leitura do manuscrito e pelos valiosos coment\'arios e sugest\~oes. Finalmente, o autor agradece a dois \'arbitros an\^onimos pela revis\~ao do manuscrito e pelas sugest\~oes fornecidas (principalmente com respeito ao idioma Portugu\^es) visando o aprimoramento deste trabalho.


\end{document}